\documentclass[useAMS,usenatbib]{mn2e}
\usepackage{graphicx}
\usepackage{natbib}
\title[MOST photometry of five young stars]{Analysis of {\it MOST} light curves of 
five young stars in Taurus-Auriga and Lupus~3 Star Forming Regions 
\thanks{Based on data from 
the MOST satellite, a Canadian Space Agency mission, jointly operated 
by Dynacon Inc., the University of Toronto Institute of Aerospace 
Studies, and the University of British Columbia, with the assistance
of the University of Vienna.}}

\author[M. Siwak et al.]
{Michal Siwak$^{1,2}$\thanks{E-mail: siwak@nac.oa.uj.edu.pl},
Slavek M.\ Rucinski$^1$,
Jaymie M.\ Matthews$^3$,
Rainer Kuschnig$^{3,7}$,
\newauthor
David B.\ Guenther$^4$,
Anthony F.\ J.\ Moffat$^5$,
Dimitar Sasselov$^6$,
Werner W.\ Weiss$^7$\\
$^1$Department of Astronomy and Astrophysics,
University of Toronto, 50 St.\ George St., Toronto,
Ontario, M5S~3H4, Canada\\
$^2$Mount Suhora Astronomical Observatory, Cracov Pedagogical University,
ul.\ Podchorazych 2, 30-084 Krakow, Poland\\
$^3$Department of Physics \& Astronomy, University of
British Columbia, 6224 Agricultural Road, Vancouver, B.C., V6T~1Z1, Canada\\
$^4$Institute for Computational Astrophysics,
Department of Astronomy and Physics,
Saint Marys University, \\
Halifax, N.S., B3H~3C3, Canada\\
$^5$D\'{e}partment de Physique, Universit\'{e}
de Montr\'{e}al, C.P.6128, Succursale: Centre-Ville,
Montr\'{e}al, QC, H3C~3J7, Canada\\
$^6$Harvard-Smithsonian Center for Astrophysics,
60 Garden Street, Cambridge, MA 02138, USA\\
$^7$Institut f\"{u}r Astronomie, Universit\"{a}t Wien,
T\"{u}rkenschanzstrasse 17, A-1180 Wien, Austria\\
}

\date{Accepted 2011 March 9;  Received 2011 March 7 ; in original form 20 January 2011}

\pubyear{2011}

\begin{document}
\label{firstpage}
\maketitle

\begin{abstract}
Continuous photometric observations 
of five young stars obtained by the {\it MOST\/} satellite in 2009 
and 2010 in the Taurus and Lupus star formation regions are presented.
Using light curve modelling
under the assumption of internal invariability of spots,
we obtained small values of the solar-type differential-rotation 
parameter ($k=0.0005-0.009$) for three spotted weak-line T~Tau stars, 
V410~Tau, V987~Tau and Lupus~3-14; for another spotted WTTS, 
Lupus~3-48, the data are consistent with a rigidly rotating 
surface ($k=0$). Three flares of similar rise 
(4~min~30~sec) and decay (1~h~45~min) times were detected
in the light curve of Lupus~3-14. 
The brightness of the classical T~Tau star RY~Tau 
continuously decreased over 3 weeks of its observations
with a variable modulation not showing any obvious periodic signal.
\end{abstract}


\section{Introduction}
\label{intro}

The Canadian {\it MOST\/} satellite, designated to obtain high precision 
photometry of bright stars, offers a unique opportunity to achieve 
good-quality, continuous light curves of fainter objects (to about 11~mag) 
with coverage extending to dozens of days. 
Among other applications, such extend time monitoring permits 
studies of stellar surface rotation rates and 
of their differential characteristics, as well as
flares on active spotted stars \citep{ruc,cr1,w2,siwak}
In this investigation, we present recently obtained {\it MOST\/} light 
curves of five T~Tau type stars.

The targets investigated in this paper are members of two different 
Star Forming Regions (SFR): Three, V410~Tau, V987~Tau and RY~Tau 
belong to the Taurus--Auriga SFR;  
we refer the interested reader to \citet{skelly}, \citet{strassmeier} 
and \citet{petrov99} for more historical details 
about these stars.
The two Lupus targets were discovered by the {\it ROSAT} satellite 
in the Lupus~3 SFR \citep{krautter}. 
Although for dozens of stars in the Lupus~3 SFR \citet{wichmann} obtained 
high resolution spectra and determined their spectral types and projected 
rotational velocities $v\,\sin i$, the stars remain rather poorly studied, 
both photometrically and spectroscopically.

Four of the targets (V410~Tau, V987~Tau, Lupus~3-14, 
Lupus~3-48) belong to the weak-line sub-type of 
T~Tau stars (WTTS) and appear to show very similar variations, 
apparently caused by large photospheric spots. 
For all these objects, the same assumption of differentially rotating 
surfaces covered by large spots can be applied.
The fifth target, RY~Tau is a classical T~Tauri star showing 
significant (previously up to a few magnitudes) and irregular 
brightness variations described 
as {\it Type~III\/} variability \citep{herbst94}
which is most probably caused by variable
absorption/extinction events due to circumstellar dust.
Table~\ref{Tab.1} gives the key physical parameters of 
targets investigated in this paper.

The paper is organized as follows: We present 
details concerning data acquisition and reduction
in Section~\ref{obs}.
The methods used for light curve analysis are described in 
Section~\ref{analysis}. 
The results obtained for individual targets 
are discussed in Section~\ref{results} and then 
summarized in Section~\ref{summary}.

\section{Observations and data reductions}
\label{obs}

The optical system of the {\it MOST\/} satellite consists
of a Rumak-Maksutov f/6, 15~cm reflecting telescope.
The custom broad-band filter covers the spectral range of
380 -- 700~nm with effective wavelength falling close
to the Johnson $V$ band.
The pre-launch characteristics of the mission are described
by \citet{WM2003} and the initial post-launch performance
by \citet{M2004}.

The stars investigated in this paper were observed in the direct-imaging 
mode of the satellite \citep{WM2003}. 
RY~Tau, V410~Tau and V987~Tau were observed together, within one
field, between 19 October and 9 November, 2009. 
Lupus~3-14 and Lupus~3-48 
were chosen as secondary targets in the field of 
the primary {\it MOST\/} target HR~5999 and were 
observed between 18 -- 30 April, 2009.
Lupus~3-14 was observed during 
a second observing run of HR~5999 between 3 and 25 May, 2010. 
The individual exposure times were 30 or 60~s long.
In general, the targets were observed during the low stray-light 
orbital phases of {\it MOST\/} which lasted typically 40 minutes of every 103 minute 
satellite orbit, although occasionally the data acquisition was
interrupted for up to a few {\it MOST\/} orbits due 
to technical problems. 
Additionally we removed a portion of
the data for RY~Tau and V410~Tau which was affected 
by reflections in the telescope optics caused by the Moon's passage 
within $\approx 4-6~\deg$.

Aperture photometry of stars was obtained using {\it dark-corrected} 
images by means of the DAOPHOT~II package \citep{stet}. 
The {\it dark} frames were obtained by averaging a dozen {\it empty-field} 
images specifically taken during each 
observing run. In the case of stars in the Taurus 
field, a weak correlation between the star flux and the sky background 
level within each {\it MOST\/} orbit, 
most probably caused by a small photometric nonlinearity 
of the electronic system (see \citet{siwak}), was noticed and removed. 
No such correlations were observed for targets in the Lupus field. 
The light curves were also corrected by low-order polynomials for 
slow photometric trends determined from a few constant stars 
observed simultaneously with the respective targets.
As a result, we obtained good quality light curves 
with median errors of the {\it mean-orbital} 
data points expressed in normalized flux units of 0.0015 (V987~Tau), 
0.0032 (V410~Tau), 0.0014 (RY~Tau), 0.0055 (Lupus~3-14) and 0.0062
(Lupus~3-48). 
The mean-orbital errors were used in calculating 
weights during the reduced and weighed $\chi^2$ calculation. 
However, as we will stress later, the model parameter uncertainties
are not driven by random errors, but rather by the
unavoidably large ranges in assumed astrophysical 
parameters such as
the flux level for the unspotted photosphere and --
particularly -- the inclination of the rotation axis.

%--------- Table 1 - Basic data of observed targets ---------
\begin{table}
\caption{Physical parameters of targets. References: 1 -- \citet{skelly}, 
2 -- \citet{strassmeier}, 
3 -- CDS,
4 -- \citet{wichmann},
5 -- \citet{pott},
* -- estimated in this paper from the available data, 
based on {\it spectral type -- effective temperature} 
calibration of \citet{harmanec}. The uncertainties, if known,
are given in parentheses.}
\begin{tabular}{l l l l l c}
\hline\hline
star          & Spc.  &$T_{{\rm eff}}$   & log~g & $v\,\sin i$ &  $i$      \\ \hline\hline
V410~Tau$^1$  & K5    & 4500       & 4.0   & 74(3)       & 70(10)          \\
V987~Tau$^2$  & G5IV  & 5250       & 3.5   & 78(1)       & 35$^{+15}_{-5}$ \\
Lupus~3-14$^3$& K0Ve  & 5190*      & ---   & ---         &  ---            \\
Lupus~3-48$^4$& K2    & 4900*      & ---   & 49.0        &  ---            \\
RY~Tau$^5$    & G1    & 5945       & ---   & 51.6        &  25(3)          \\ \hline\hline 
\end{tabular}
\label{Tab.1}
\end{table}
%--------------------------------------------------------------------------

\section{Light curve analysis. The methods}
\label{analysis}

We used two different techniques for analysis of the four
spotted stars and for RY~Tau which showed a complex, evolving
light curve:\newline
(1)~For the light curve of RY~Tau, we used the 
Fourier transform technique for unevenly sampled data;
the results are described in Section~\ref{rytau}.\newline
(2)~For the four spotted stars, 
we used the spot model program {\it StarSpotz}, as described
in \citet{cr1} and \citet{w2}. 

Because the results for RY~Tau are rather inconclusive, 
we concentrated on determination of the differential rotation
of the four spotted stars assuming that the 
dark spots are invariable in time (except for longitudinal  
drifts caused by differential rotation). 
This assumption is supported by the results of
several Doppler imaging campaigns 
for V410~Tau and V987~Tau 
(see Sections~\ref{v410} and \ref{v987}).
As it will be shown below, their global, progressive light 
curve shape changes can be fully explained by differential 
rotation of their stellar surface. 
Any random, rapid or small scale changes of the spots, appear to be absent 
or have been averaged over duration of the {\it MOST} observations.

During the light curve modelling process, 
we assumed the following parameters as constant 
(if known, see Table~\ref{Tab.1}): the rotation axis inclination 
$i$, the stellar radius $R$ and the projected rotational velocity 
$v\,\sin i$. 
Assuming photospheric and spot temperatures  
of $T_{\rm eff} = 4500$~K and $T_{\rm spot} = 3600$~K \citep{skelly} 
for V410~Tau and $T_{\rm eff} = 5250$~K 
and $T_{\rm spot} = 3800$~K for V987~Tau 
\citep{strassmeier}, the {\it spot-to-photosphere} flux ratio 
$f = 0.077 \pm 0.015$ and $f = 0.105 \pm 0.015$ was evaluated 
for the {\it MOST\/} filter bandpass by means of the SPECTRUM programme 
\citep{gray} and Kurucz's atmosphere models \citep{kurucz}.
The same value of $f$ was assumed for all spots on a given star.
Because the {\it MOST\/} magnitudes are very close to those of Johnson $V$ 
ones \citep{siwak}, we assumed the linear limb-darkening coefficient 
$u=0.817$ (for V410~Tau) and $u=0.737$ (for V987~Tau), 
calculated for the Johnson $V$-band by \citet{dc1}.
Due to the absence of data on 
$\log\,g$ for Lupus~3-14 and Lupus~3-48, 
we assumed the identical value of $f=0.08$ and the limb 
darkening corresponding to their spectral types of
$u=0.737$ and $u=0.78$; we note that
the exact value of $u$ has a second-order effect on the
predicted light curves compared with effects caused by the assumed 
simple circular geometry of spots. 
Except for V410~Tau, the unspotted magnitudes and hence
the corresponding values of spot-free stellar fluxes ($F_u$) were 
not known for our targets; for that reason, 
we performed a grid of solutions for 
several fixed values of $F_u$ (1.0, 1.1, ..., 1.7).
In Table~\ref{Tab.2} we present the results obtained for the 
best-fitting values of $F_u$.
\newline
The adjusted parameters of the model were, for each of two spots: 
(1)~the initial moment $t$ in $hjd_1 \equiv HJD-2,455,100$ (for stars 
in the Taurus field) or $hjd_2 \equiv HJD-2,454,900$ (for stars in the 
Lupus~3 field), when the spot is facing the observer, 
(2)~the rotation period $p$ of the spot in days, 
(3)~the latitude ${\phi}$ of the spot in degrees, and 
(4)~its diameter $r$ in degrees.

During the search for the solar-type differential rotation 
coefficient $k$, the individual spot rotation periods 
$p_j$ were assumed to depend on the stellar latitude $\phi_j$, 
the rotational period at the equator $P_{eq}$ (in days) and 
on the differential-rotation parameter $k$ through 
the solar-type rotation law:
\begin{equation}
p_{j}({\phi_j})=P_{eq}/(1 - k \sin^{2}{\phi_j}),
\end{equation}
where $j=1,2$ denote the individual spots.

The search procedure for $k$ in the cases of V410~Tau and V987~Tau, 
consisted of two steps:
first, for the assumed stellar radius $R$ (in solar units) we searched for 
the value of $P_{eq}$ best corresponding to the observed $v\, \sin i$; 
then in the second step, we searched for $k$ which would return 
the smallest value of the reduced and weighted ${\chi}^2$.
A similar procedure was applied during the Lupus~3-48 light curve modelling, 
but because the values of 
$R$ and $i$ are unknown, we assumed a wide range
of inclinations within $i=60 \pm 20~\deg$ and 
then adjusted $R$ to match the observed value of $v\,\sin i$.
As no parameters are known for Lupus~3-14, we limited our 
investigation to quantitative estimations of $P_{eq}$ and $k$ only.

Although formal errors of the parameters can be calculated from the search 
matrix returned during MCMC computations, their values turn out to be 
unrealistically small. 
We resolved this problem by considering extreme acceptable values 
of the inclination in driving the uncertainties in the model results -- 
this is the parameter whose uncertainty has the strongest 
impact on the obtained results. 
Therefore in Table~\ref{Tab.2}, we present the range of possible values from 
light curve models obtained for the assumed lower and highest possible values 
of $i$.

% ----------------------- Fig.1 the light curve ---------------------
\begin{figure*}
\includegraphics[height=150mm,angle=-90]{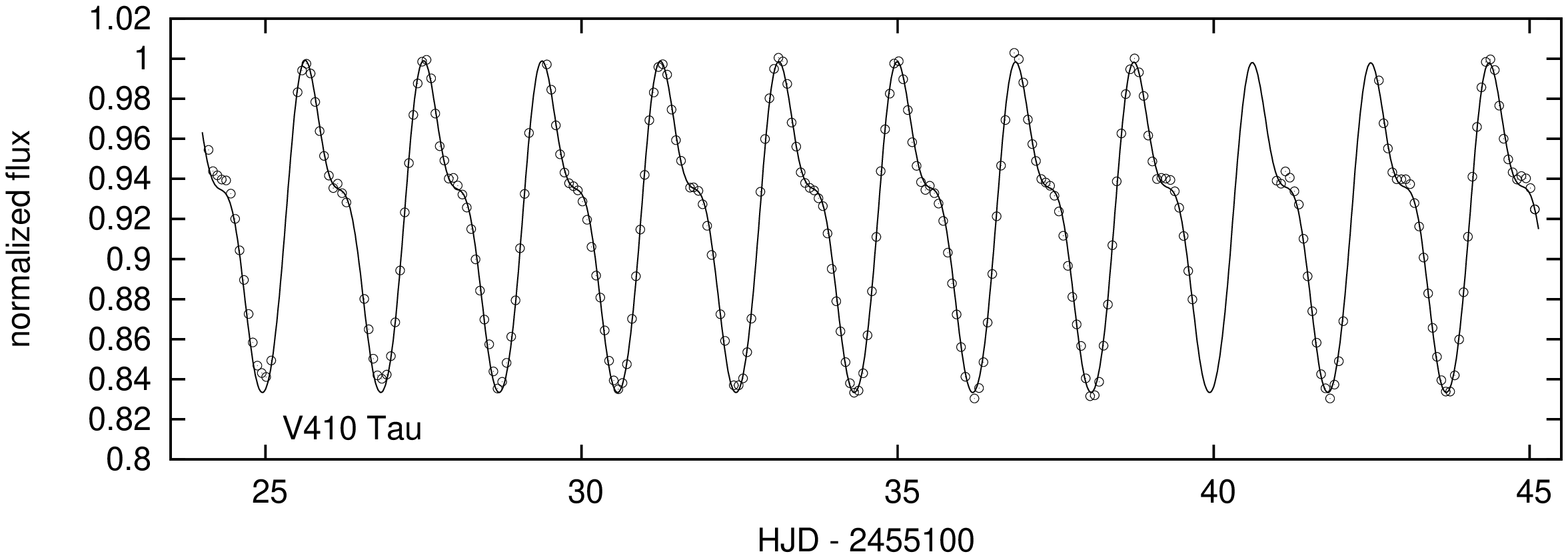}
\includegraphics[height=150mm,angle=-90]{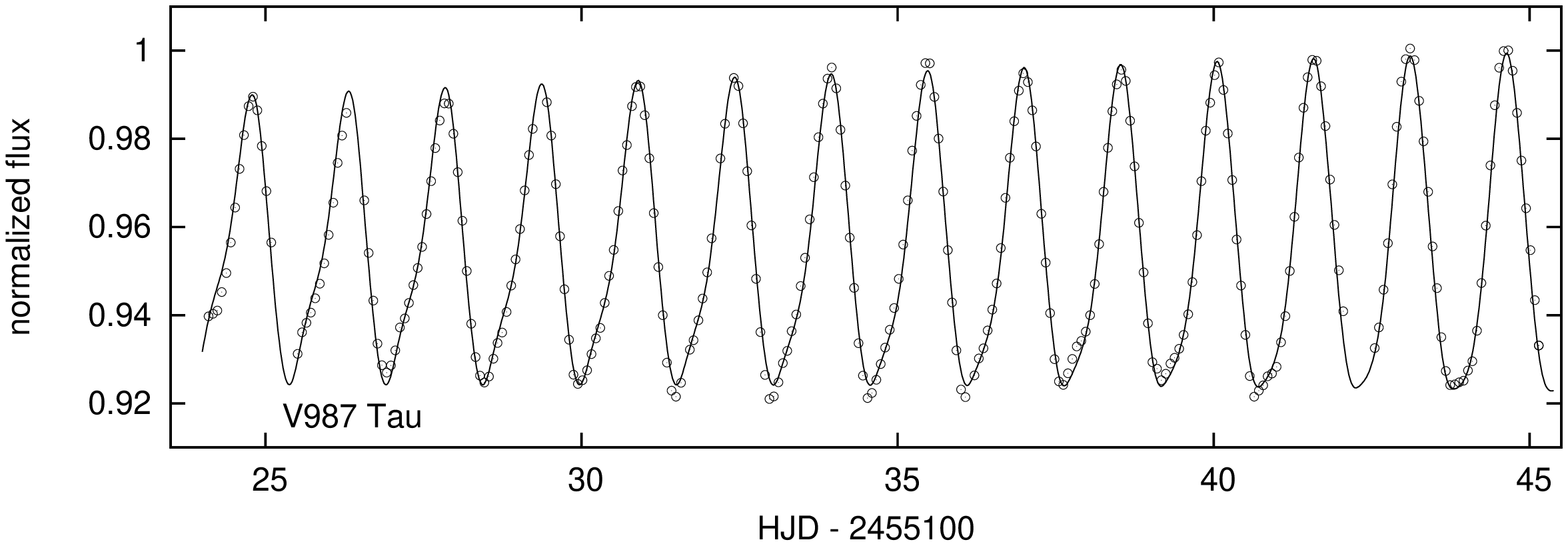} 
\includegraphics[height=150mm,angle=-90]{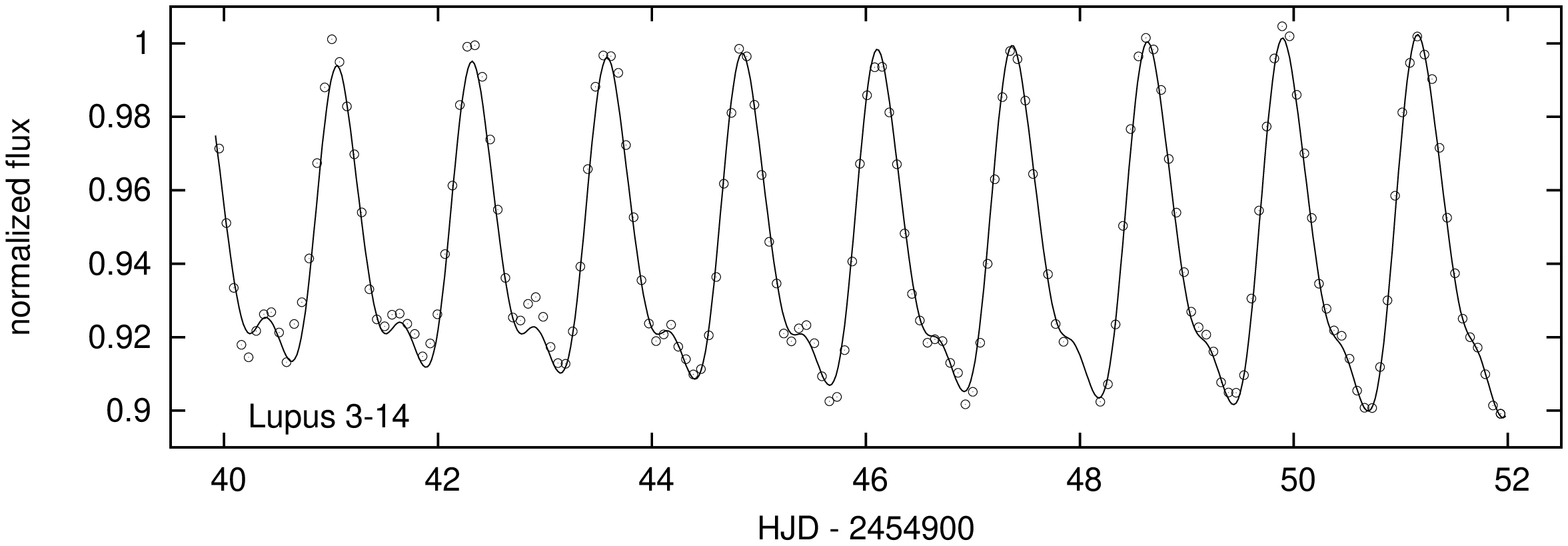} 
\includegraphics[height=150mm,angle=-90]{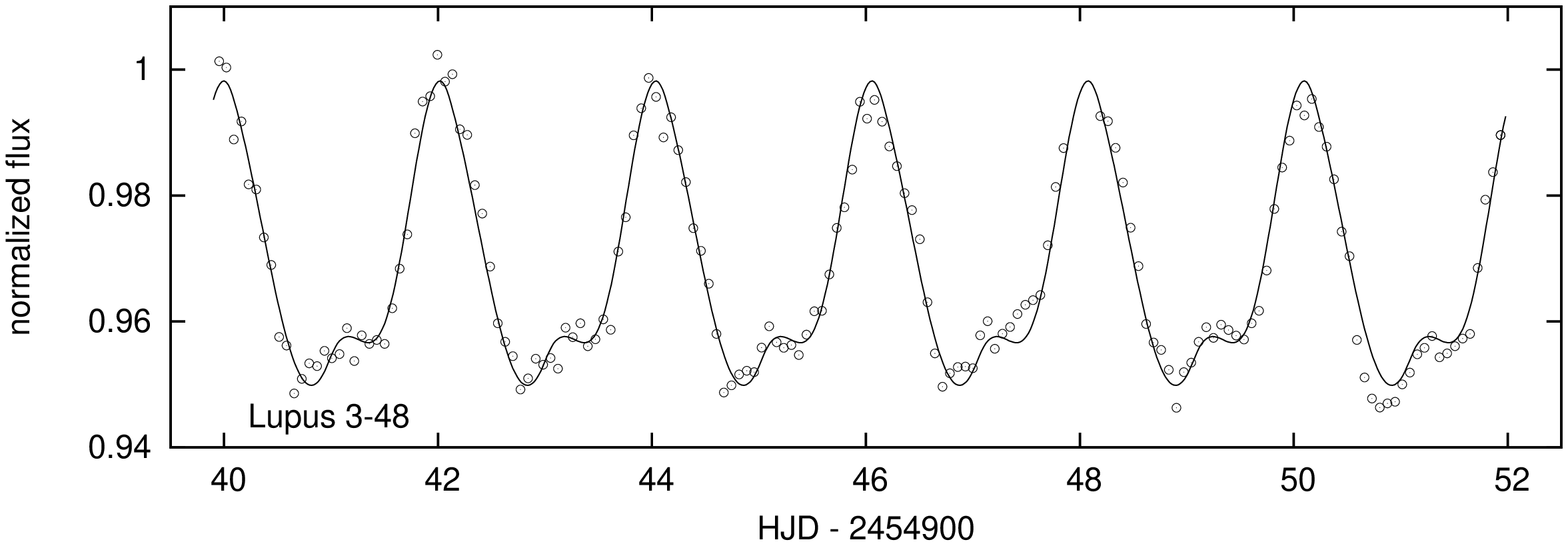}
\caption{The comparison between the {\it mean-orbital} {\it MOST\/} data 
(circles) and the synthetic light curves. The flares detected 
in the Lupus 3-14 light curve, were removed during computations.
The light curves of the first three stars show systematic 
changes which can be interpreted as a sign of a differential rotation 
of their stellar surfaces, 
fully confirmed by the light curve modelling.
A solid-body rotation model provides a very good fit to the data 
of the fourth star.}
\label{Fig.1}
\end{figure*}
%----------------------------------------------------------------------

% ---------------------Table 2 - results of the light curve modelling -------------------------------------------
\begin{table*}
\centering
\begin{minipage}{170mm}
 \caption{The light curve models for the four WTT stars, 
obtained for the assumed inclinations $i$. 
The ranges of the derived parameters given in parentheses 
correspond to the extreme possible values of $i$, 
as given in Table~\ref{Tab.1}, 
with the exception of the targets from the Lupus~3 SFR, 
where we assumed $i = 60 \pm 20~deg$.
 \newline a -- assumed as constant during modelling, b -- determined using the constraints 
given by $R$ and $v \sin i$, c -- calculated using the Eq.(1), d -- see Section~\ref{lupus3-48} for details.
}    
\begin{tabular} {@{}lllll@{}}
 \hline\hline
 Star:              & V410~Tau                & V987~Tau                    & Lupus 3-14                & Lupus 3-48 \\ \hline\hline
 $F_u$              & 1.55$^a$                &  1.1$^a$                    &  1.1$^a$                  &  1.0$^a$ \\
 $i~[^\circ]$       &   70~~(60 -- 80)$^a$    &   35~~(30 -- 50)$^a$        &  60 (40 -- 80)$^a$        &   60~~(40 -- 80)$^a$ \\
 $v \sin i$ [km/s]  &   74$^a$                &   78$^a$                    & ---                       &   49$^a$ \\        
 $R~[R_{\odot}]$    & 2.9 (3.2 -- 2.8)        & 4.1 (4.7 -- 3.1)            & ---                       & 2.3 (3.0 -- 2.0)  \\
 $P_{eq}$~[d]       & 1.87197$^b$             &1.5215 (1.5220 -- 1.5210)$^b$& 1.257 (1.259 -- 1.256)    & ---      \\      
 $k$                & 0.0005(1)$^b$           &0.0064 (0.0063 -- 0.0066)$^b$& 0.009 (0.008 -- 0.01)     & ---      \\ \hline
 $p_1$~[d]          & 1.87301$^c$          &1.52206 (1.52240 -- 1.52243)$^c$& 1.2681 (1.2683 -- 1.2684)$^c$& 2.01973 (2.01971 -- 2.01964)\\
 $t_1$~[hjd$_{1,2}$]&24.952 (24.950 -- 24.970)& 24.452~~(24.449 -- 24.450)  & 40.241~~(40.241 -- 40.260)& 40.780~~(40.780 -- 40.768)   \\
 ${\phi}_1~[^\circ]$& 75.4~~~~(76.0 -- 71.6)  & 13.9~~~~~(11.8 -- 22.0)     & 80.1~~~~~(73.2 -- 81.7)   & 60.4~~~~~(41.7 -- 72.0)  \\
 $r_1~[^\circ]$     & 84.8~~~~(77.2 -- 90.0)  & 13.7~~~~~(15.6 -- 11.0)     & 33.4~~~~~(24.7 -- 48.3)   & 13.4~~~~~(11.8 -- 22.0)  \\ \hline
 $p_2$~[d]          & 1.87266$^c$          &1.53052 (1.53053 -- 1.53077)$^c$& 1.2605 (1.2596 -- 1.2613)$^c$& 2.01973 (2.01971 -- 2.01964)\\
 $t_2$~[hjd$_{1,2}$]&25.977 (25.982 -- 25.974)& 25.378~~(25.373 -- 25.382)  & 40.715~~(40.722 -- 40.714)& 41.472~~(41.473 -- 41.455)  \\
 ${\phi}_2~[^\circ]$& -52.3~~~(-48.9 -- -58.1)& 73.6~~~~~(70.2 -- 79.5)     & 33.7~~~~~(13.4 -- 40.5)   & ~7.5~~~~~(0.4 -- 36.8)     \\
 $r_2~[^\circ]$     & 44.6~~~~(48.0 -- 49.8)  & 24.1~~~~~(23.3 -- 29.0)     & 15.5~~~~~(19.1 -- 17.0)   & 10.5~~~~~(14.3 -- 10.9)   \\ \hline
 ${\chi}^2_{red,w}$ & 1.3778                  & 2.6903                      & 0.3916$^d$                    & 0.2391$^d$   \\ \hline\hline
\end{tabular}
\label{Tab.2}
\end{minipage}
\end{table*}
% --------------------------------------------------------------------------

\section{Results for individual targets}
\label{results}

\subsection{V410~Tau}
\label{v410}
Doppler tomographic studies \citep{joncour94a, hatzes, rice}
and spectropolarimetric observations of V410~Tau \citep{skelly}, 
revealed one large cold spot at high latitudes, 
and one or a few smaller cold spots at moderate or low latitudes.
From Doppler maps obtained in the time interval of one year 
\citet{rice} derived a small value of the solar-type differential 
rotation of $k=0.001$, however, the analysis of \citet{skelly} 
indicated a rigidly rotating stellar surface for the star.

In accordance with previous results of Doppler tomographic studies, 
we started computations assuming two dark spots located near 
to the poles and separated in longitude by nearly 180~degrees. 
The final spot configuration found by the model (Table~\ref{Tab.2}) 
is in good accordance with the most recent Doppler images obtained 
by \citet{skelly}. 
These authors obtained the rotational period of 1.87197~d, but because 
at that time the star was probably considerably more uniformly spotted 
(as inferred from the small amplitude of the photometric variability of 
only 0.04~mag compared with 0.2 mag which we observed 10 months later) 
this value may be close to the equatorial rotational period 
of the star. 

Assuming the above value of the photometric period as equal to $P_{eq}$, 
we obtained the observed $v\,\sin i = 74~km/s$ by setting $R$ 
equal to 2.91~$R_{\odot}$. 
The best model resulted in a small amount of differential rotation 
for this star, $k=0.0005(1)$ (Table~\ref{Tab.2}). 
A solid--body model provided a slightly 
worse fit, with $\chi_{red,w}^{2}$ by 4\% higher.
The computations were obtained for 
the unspotted flux $F_u=1.55$, which is the average 
of 1.4 and 1.7, as obtained from the comparison 
between the most probable value 
of the stellar maximum brightness during 
the {\it MOST\/} observations ($V_{max} \approx 10.8~mag$), 
and the two values of unspotted magnitude (corresponding
to $F_u$), estimated by \citet{petrov} at $V_u=10.44$ 
and by \citet{grankin} at $V_u=10.236$.

\citet{fernandez} summarized several historical observations
of flares, and reported detection
of nine flares during 11~days of V410~Tau monitoring in November 2001.
The flare amplitudes in the Johnson $V$-filter were 
rather moderate, 0.02 -- 0.1~mag,
but one flare reached almost 1~mag.
The flares did show a tendency to appear in the light curve minimum,
when the most active spotted region was directed to the observer. 
A careful inspection of the {\it MOST\/} light curve, obtained
during the 22~day long run, did not reveal any flares larger than 0.01 
magnitudes and of similar duration to those
observed by \citet{fernandez}. 
We stress that our result may be affected by interruptions 
of {\it MOST\/} observations during the high stray-light phases, which
altogether lasted about 60\% of the total coverage time.

\subsection{V987~Tau (HDE 283572)}
\label{v987}
The pre-main sequence nature of V987~Tau (HDE~283572), one of 
the brightest known weak-line T~Tauri stars was established by \citet{walter}. 
Two Doppler imaging sessions \citep{joncour94b, strassmeier}, 
revealed a large cold spot in the polar region. 

To retain consistency with this result and 
to explain the systematic light curve changes, we assumed two dark 
spots: one close to the pole and directed to the observer, 
and a second at low latitudes.
For all considered values of the spot-free flux
$F_u$ (see Section~\ref{analysis}), 
the best model yielded almost the same value of the differential 
rotation parameter $k$, about 0.0064.
In Table~\ref{Tab.2}, we present the values obtained 
for $F_u$=1.1, giving the smallest value of $\chi^{2}_{red,w}$. 
According to \citet{johns-krull}, the star seems to rotate 
as a solid-body, but with this assumption we obtained 2.5 times 
larger $\chi^{2}_{red,w}$ than for $k \ne 0$.
We also note that from the {\it MOST\/} observations we obtained 
a value of the rotation period that is about 40 min shorter than 
1.5495~d, as obtained by \citet{strassmeier}.

\subsection{Lupus~3-14 (RXJ1605.8-3905)}
\label{lupus3-14}
In this investigation, we limited ourselves to light curve modelling 
of the 2009 data only -- the 2010 data show very simple, sine-like 
variations and contain little information about differential rotation. 
It is worth noting that both the 2009 and 2010 light curves 
showed regular trends by about 0.03~mag, occurring on a timescale 
of about 20 days. 
As they appear to be similar to ellipsoidal variations observed 
in binary stars, this may suggest that Lupus~3-14 is a binary star.

As none of the stellar parameters 
($R$, $\log\,g$, $v \sin i$, $i$) is known 
for Lupus~3-14, we arbitrarily assumed the inclination 
$i=60$ with a very wide range of $\pm 20\deg$, 
i.e.\ $i = 40 - 80 \deg$. 
Under the assumption of a rigidly rotating stellar surface, 
the 2009 data gave $p_{1,2}=1.2649$~d, but 
a rather poor fit, mostly due to the 
constantly changing shape of the light curve. 
Considering different values of $P_{eq}$ and $k$, the best fits resulted 
in considerably smaller $\chi^2_{red,w}$ than in the case of the solid body 
rotational model (Tab.~\ref{Tab.2}).
Two dark spots, one located near the pole, which is directed to the observer, 
and one at moderate 
latitudes sufficed to explain the systematic light changes in great detail. 
The results are rather independent of $F_u$ (although the value of $F_u=1.1$ yielded 
the best fit to the data); we feel that determination of this parameter may 
require long term observations, like these started by \citet{grankin08}. 
As long as the exact value of $F_u$ remains unknown, the derived  
spots radii $r_i$, especially of high--latitude spots (visible through the 
whole rotation), will remain affected due to the strong correlation 
between these parameters. 

During the 2009 and 2010 observing runs of Lupus~3-14, {\it MOST\/} 
detected three flares.
The outburst amplitudes of the three flares are similar to within about 
10\% in continuum flux units, although the amplitude of the 2009 flare is 
slightly smaller, possibly due to the two-times longer (60~s) exposure 
used. 
From the data of the flare observed in 2009 and from the second flare 
observed in 2010, we estimated the rise time rate (from the continuum 
to maximum) at 4~m~30~s. 
The decline rate of the first 2010 flare was determined to be about 
1~h~45~m, which is comparable with the orbital period of the {\it MOST\/} 
satellite and hence only roughly estimated.
The same decline rate is most probably observed for all flares, as after 
each outburst event the observations gathered during the next {\it MOST\/} 
orbit had levels coincident with the unperturbed level.

\subsection{Lupus 3-48 (RXJ1608.9-3905)}
\label{lupus3-48}
Similary as for the previous target, 
during the computations we arbitrarily assumed $i=60\pm20~deg$. 
With this assumption, $P_{eq}=2.02~d$ and $R=2.25~R_{\odot}$ reproduce 
the observed $v\,\sin i$ = 49~km/s and (with $F_u = 1.0$) provide the smallest $\chi^2_{red,w}$. 
The model predicted one spot at high latitudes and a second spot close 
to the stellar equator.
Unfortunately, because of the small amplitude of the light curve and 
relatively large errors of individual data points (probably caused by 
the intrinsic variability of the star and visible in the light curve minima), 
we were unable to find 
whether the star rotates differentially; in fact, the solution with $k=0$ 
suffices to explain the available data (Tab.~\ref{Tab.2}, Fig.~\ref{Fig.1}).

We note that due to the large errors of the {\it mean-orbital}
data points of both stars from the Lupus field (caused by 
spurious and non-removed variability of the
star flux within individual {\it MOST\/} orbits, see Section~\ref{obs}), 
the values of ${\chi}^2_{red,w}$ given in Table~\ref{Tab.2} are considerably smaller than unity. The
respective values of ${\chi}^2_{red,w}$ 
can be brought to unity assuming median 
orbital errors of 0.0034 and 0.0028 for Lupus~3-14 
and Lupus~3-48, respectively.

\subsection{RY~Tau}
\label{rytau}
RY~Tau is a prototype of the {\it Type~III\/} T~Tauri variability 
\citep{herbst94}, which is most likely caused by large dust clouds 
forced to co-rotation by magnetic fields of the star. 
The occultations are irregular and colour independent \citep{petrov99}.

\begin{figure}
\includegraphics[width=55mm,angle=-90]{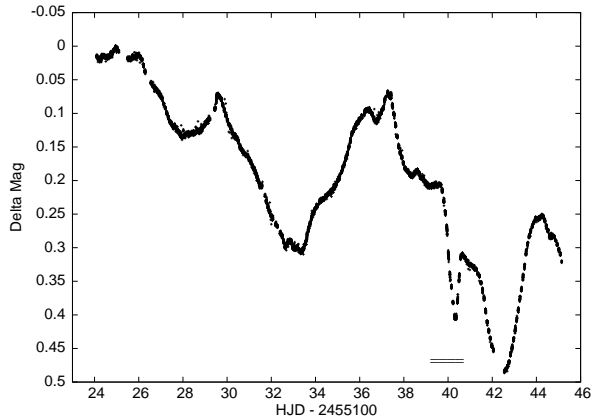}
\caption{The light curve of RY~Tau obtained by {\it MOST\/} in magnitude 
scale.
A section of the data marked by a horizontal bar
at $hjd_1 = 39.6-40.6$ was obtained during 
close Moon passage at an angular distance of 4--6 degrees 
from the target which has been excluded in the 
Fourier analysis.}
\label{Fig.2}
\end{figure}

\begin{figure}
\includegraphics[width=80mm]{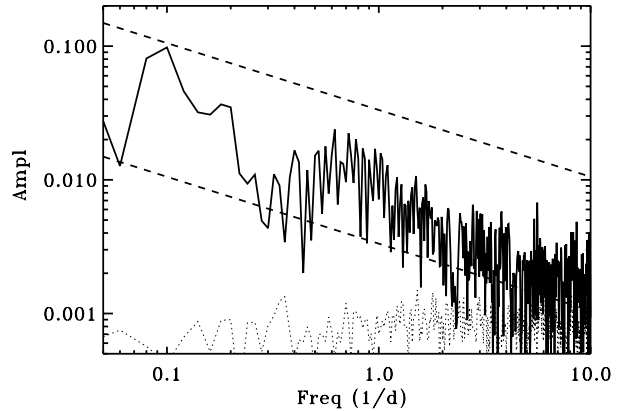}
\caption{The Fourier transform (thick line) in log-log scale of the RY Tau light curve, 
         detrended by a linear fit. The amplitude errors are shown as small dots.}
\label{Fig.3}
\end{figure}

The {\it MOST\/} data indicate that the brightness of the star was falling 
during the run with two superimposed, short time-scale brightness decreases 
which occurred at $hjd_1$ = 28 and 33 (Figure~\ref{Fig.2}). 
Similar brightness drops appeared at $hjd_1$ = 38 and 42.5, but they 
were seriously influenced by the close Moon passage which caused 
a false symmetrical brightness dip at $hjd_1$ = 39.6 -- 40.6 (this is marked 
by a horizontal bar in the figure).
If the former brightness drops were caused by a transit 
of the same dusty clouds, then their orbital period would 
be about 9 -- 10~days.
Indeed, the Fourier spectrum of the data corrected for the continuous 
downward trend in brightness (by a linear fit) reveals many statistically 
significant peaks with that at 0.1~c/d related to such a likely 
periodicity (Fig.~\ref{Fig.3}). However, expected orbital
periods of disk condensations are longer: 
For the inner disk radius (0.3~AU) and the 
stellar mass of $1.69-2.00\,M_{\odot}$ \citep{schegerer}, 
the Keplerian orbital period is about 45~days.
Such a long duration would correspond to 
an upper limit for the duration of the {\it MOST\/} runs and did not appear 
feasible for the operation reasons of the satellite.
For the period of 9--10 d, the expected distance from the
star would be about 22~$R_{\odot}$, a distance 
which is not excluded for a model of magnetic field lines 
enforcing co-rotation with the central star \citep{petrov90,herbst94}.

% ----------------------- Fig.2 the light curve ---------------------
%----------------------------------------------------------------------

% ------------------- Fig.3 the fourier transform ---------------------
%----------------------------------------------------------------------

\section{Summary}
\label{summary}

Continuous {\it MOST\/} satellite observations of four 
weak-line T~Tauri stars led to derivation of  
small values of the solar-type, differential-rotation 
parameter $k$.  
In three cases, $k$ was found to be equal within the errors 0.0005(1), 0.0064(2), and 0.009(1) 
for V410~Tau, V987~Tau and Lupus~3-14, respectively. 
In all these three cases, the differential rotation provided a significantly 
better goodness-of-fit parameter ($\chi^2_{red,w}$) and satisfactorily explained
the systematic evolution of the light curves (Fig.~\ref{Fig.1}). 
In the fourth case of Lupus~3-48, the small amplitude of the light variations 
and rather large errors precluded derivation of the parameter $k$, but 
the solid-body rotation model satisfactory explains the available data.
In all our solutions we obtained spot distributions in good accord with 
the results of previous Doppler imaging studies of the targets, i.e.\ 
with one large spot, located near a pole (possibly a remnant of 
a dipole magnetic-field topology from the classical T~Tauri phase), 
and a second spot at low latitudes.\newline
In this investigation we limited ourselves to test the solar-type and 
the rigid body differential rotation laws; with a more reliable model of spot 
distribution, we plan to perform a test of the anti-solar and other rotation 
laws.

Although all the investigated stars are known to be chromospherically
active, the {\it MOST\/} data have shown flares only in 
the case of Lupus~3-14. 
For all the three large flares on this star the rise (4~m~30~s) and decay 
(1~h~45~m) times were similar.

An even longer observing run may 
shed more light on the short time scale 
phenomena occurring in the RY~Tau light curve; the suggested orbital period 
(9 -- 10~day) of small dusty clouds is only a conjectural conclusion.

\section*{Acknowledgments}

MS acknowledges the Canadian Space Agency Post-Doctoral position
grant to SMR within the framework of the Space Science Enhancement Program.
The Natural Sciences and Engineering Research Council of
Canada supports the research of DBG, JMM, AFJM, and SMR.
Additional support for AFJM comes from FQRNT (Qu\'ebec).
RK is supported by the Canadian
Space Agency and WWW is supported by the Austrian Space
Agency and the Austrian Science Fund.

Particular thanks are due to Mr. Bryce Croll, for preparing a special 
version of the {\it StarSpotz} light curve modelling programme, and to 
the anonymous referee for useful comments and suggestions.

This research has made use of the SIMBAD database,
operated at CDS, Strasbourg, France and NASA's Astrophysics
Data System (ADS) Bibliographic Services.

\end{document}